\documentclass[aps,prl,preprint,nopacs,superscriptaddress]{revtex4}
\usepackage{graphicx}
\usepackage{verbatim}
\usepackage{mathrsfs}
\usepackage{epstopdf}
\usepackage{amsmath,amsfonts,amssymb}
\usepackage{graphicx}

\def\3{2.8in}    
\def\2{2.5in}
\def\4{3.0in}

\def \beq {\begin{equation}}
\def \eeq {\end{equation}}
\begin{document}

\title{Discovery of Weyl semimetal NbAs}

\author{Su-Yang Xu$^*$}
\affiliation {Laboratory for Topological Quantum Matter and Spectroscopy (B7), Department of Physics, Princeton University, Princeton, New Jersey 08544, USA}
\affiliation{Princeton Center for Complex Materials, Princeton Institute for Science and Technology of Materials, Princeton University, Princeton, New Jersey 08544, USA}

\author{Nasser Alidoust$^*$}\affiliation {Laboratory for Topological Quantum Matter and Spectroscopy (B7), Department of Physics, Princeton University, Princeton, New Jersey 08544, USA}
\affiliation{Princeton Center for Complex Materials, Princeton Institute for Science and Technology of Materials, Princeton University, Princeton, New Jersey 08544, USA}
\author{Ilya Belopolski\footnote{These authors contributed equally to this work.}}\affiliation {Laboratory for Topological Quantum Matter and Spectroscopy (B7), Department of Physics, Princeton University, Princeton, New Jersey 08544, USA}
\affiliation{Princeton Center for Complex Materials, Princeton Institute for Science and Technology of Materials, Princeton University, Princeton, New Jersey 08544, USA}

\author{Chenglong Zhang}\affiliation{International Center for Quantum Materials, School of Physics, Peking University, China}

\author{Guang Bian}\affiliation {Laboratory for Topological Quantum Matter and Spectroscopy (B7), Department of Physics, Princeton University, Princeton, New Jersey 08544, USA}

\author{Tay-Rong Chang}
\affiliation{Department of Physics, National Tsing Hua University, Hsinchu 30013, Taiwan}
\affiliation {Laboratory for Topological Quantum Matter and Spectroscopy (B7), Department of Physics, Princeton University, Princeton, New Jersey 08544, USA}

\author{Hao Zheng}\affiliation {Laboratory for Topological Quantum Matter and Spectroscopy (B7), Department of Physics, Princeton University, Princeton, New Jersey 08544, USA}
\author{Vladimir Strokov}\affiliation {Paul Scherrer Institute, Swiss Light Source, CH-5232 Villigen PSI, Switzerland}
\author{Daniel S. Sanchez}\affiliation {Laboratory for Topological Quantum Matter and Spectroscopy (B7), Department of Physics, Princeton University, Princeton, New Jersey 08544, USA}

\author{Guoqing Chang}
\affiliation{Centre for Advanced 2D Materials and Graphene Research Centre National University of Singapore, 6 Science Drive 2, Singapore 117546}
\affiliation{Department of Physics, National University of Singapore, 2 Science Drive 3, Singapore 117542}

\author{Zhujun Yuan}\affiliation{International Center for Quantum Materials, School of Physics, Peking University, China}

\author{Daixiang Mou}\affiliation{Division of Materials Science and Engineering, Ames Laboratory, U.S. DOE, Ames, Iowa 50011, USA}\affiliation{Department of Physics and Astronomy, Iowa State University, Ames, Iowa 50011, USA}
\author{Yun Wu}\affiliation{Division of Materials Science and Engineering, Ames Laboratory, U.S. DOE, Ames, Iowa 50011, USA}\affiliation{Department of Physics and Astronomy, Iowa State University, Ames, Iowa 50011, USA}
\author{Lunan Huang}\affiliation{Division of Materials Science and Engineering, Ames Laboratory, U.S. DOE, Ames, Iowa 50011, USA}\affiliation{Department of Physics and Astronomy, Iowa State University, Ames, Iowa 50011, USA}

\author{Chi-Cheng Lee}
\affiliation{Centre for Advanced 2D Materials and Graphene Research Centre National University of Singapore, 6 Science Drive 2, Singapore 117546}
\affiliation{Department of Physics, National University of Singapore, 2 Science Drive 3, Singapore 117542}

\author{Shin-Ming Huang}
\affiliation{Centre for Advanced 2D Materials and Graphene Research Centre National University of Singapore, 6 Science Drive 2, Singapore 117546}
\affiliation{Department of Physics, National University of Singapore, 2 Science Drive 3, Singapore 117542}


\author{BaoKai Wang}
\affiliation{Centre for Advanced 2D Materials and Graphene Research Centre National University of Singapore, 6 Science Drive 2, Singapore 117546}
\affiliation{Department of Physics, National University of Singapore, 2 Science Drive 3, Singapore 117542}
\affiliation{Department of Physics, Northeastern University, Boston, Massachusetts 02115, USA}
\author{Arun Bansil}
\affiliation{Department of Physics, Northeastern University, Boston, Massachusetts 02115, USA}
\author{Horng-Tay Jeng}
\affiliation{Department of Physics, National Tsing Hua University, Hsinchu 30013, Taiwan}
\affiliation{Institute of Physics, Academia Sinica, Taipei 11529, Taiwan}

\author{Titus Neupert}\affiliation {Princeton Center for Theoretical Science, Princeton University, Princeton, New Jersey 08544, USA}
\author{Adam Kaminski}\affiliation{Division of Materials Science and Engineering, Ames Laboratory, U.S. DOE, Ames, Iowa 50011, USA}\affiliation{Department of Physics and Astronomy, Iowa State University, Ames, Iowa 50011, USA}

\author{Hsin Lin}
\affiliation{Centre for Advanced 2D Materials and Graphene Research Centre National University of Singapore, 6 Science Drive 2, Singapore 117546}
\affiliation{Department of Physics, National University of Singapore, 2 Science Drive 3, Singapore 117542}

\author{Shuang Jia}
\affiliation{International Center for Quantum Materials, School of Physics, Peking University, China}\affiliation{Collaborative Innovation Center of Quantum Matter, Beijing,100871, China}

\author{M. Zahid Hasan}
\affiliation {Laboratory for Topological Quantum Matter and Spectroscopy (B7), Department of Physics, Princeton University, Princeton, New Jersey 08544, USA}\affiliation{Princeton Center for Complex Materials, Princeton Institute for Science and Technology of Materials, Princeton University, Princeton, New Jersey 08544, USA}


\begin{abstract}
Three types of Fermions play a fundamental role in our understanding of nature: Dirac, Majorana, and Weyl. While Dirac fermions are known, the latter two have not been observed as any fundamental particle in high energy physics and have emerged as a much-sought-out treasure in condensed matter physics. A Weyl semimetal is a novel crystal whose low-energy electronic excitations behave as Weyl fermions. It has received worldwide interest and is believed to open the next era of condensed matter physics after graphene and three-dimensional topological insulators. However, experimental research has been held back because Weyl semimetals are extremely rare in nature. Here, we present the experimental discovery of the Weyl semimetal state in niobium arsenide (NbAs). Utilizing the combination of soft X-ray and ultraviolet photoemission spectroscopy, we systematically study both the surface and bulk electronic structure of NbAs. We experimentally observe both the Weyl cones and Weyl nodes in the bulk and the Fermi arcs on the surface of this system. Our ARPES data, in agreement with our previous theoretical prediction and present band structure calculations, provide conclusive evidence for the topological Weyl semimetal state in NbAs. Our discovery not only paves the way for the many fascinating topological quantum phenomena predicted in Weyl semimetals, but also establishes a new cornerstone in the correspondence between high-energy and condensed matter physics.
\end{abstract}
\date{\today}
\maketitle


Weyl semimetals have received significant attention in recent years because they extend the classification of symmetry-protected topological phases beyond insulators, demonstrate unusual transport phenomena and provide an emergent condensed matter realization of Weyl fermions, which do not exist as fundamental particles in the standard model \cite{Weyl, Balents_viewpoint, Wilczek, Ashvin_Review, Haldane, TI_book_2014, Wan2011, Murakami2007, Burkov2011, HgCrSe, Thallium, HgCdTe, Vanderbilt, Suyang, Nielsen1983, Chiral, Aji2012, Nonlocal, Carbotte2013, Nielsen1983b, Franz2013, Hasan_Na3Bi}. Such kind of topologically nontrivial semimetals are believed to open a new era in condensed matter physics. In contrast to topological insulators where only the surface states are interesting, a Weyl semimetal features unusual band structure in the bulk and on the surface, leading to novel phenomena and potential applications. This opens up unparalleled research opportunities, where both bulk and surface sensitive experimental probes can measure the topological nature and detect quantum phenomena. In the bulk, a Weyl semimetal has a band structure with band crossings, Weyl nodes, which are associated with definite chiral charges. Unlike the two-dimensional Dirac points in graphene, the surface state Dirac point of a three-dimensional topological insulator or the three-dimensional Dirac points in the bulk of a Dirac semimetal, the degeneracy associated with a Weyl node does not require any symmetry for its protection other than the translation symmetry of the crystal lattice. The low-energy quasiparticle excitations of a Weyl semimetal are chiral fermions described by the Weyl equation, well-known in high energy physics, which gives rise to a condensed matter analog of the chiral anomaly associated with a negative magnetoresistance in transport \cite{Nielsen1983, Chiral,Aji2012,Nonlocal,Carbotte2013,Franz2013}. On the surface, the nontrivial topology guarantees the existence of surface states in the form of ``Fermi arcs'', which are open curves that connect the bulk Weyl nodes. The Fermi arcs on the surface of a Weyl semimetal are by themselves of great interest, because Fermi surfaces have to be closed contours in any purely two-dimensional band structure. Moreover, they are further interesting because they lead to novel spin polarization textures \cite{Ojanen}, unusual quantum interference effects in tunneling spectroscopies \cite{Hosur}, and a new type of quantum oscillation \cite{Ashvin2}, where electrons move in real space between different surfaces of a bulk sample when executing a constant-energy orbit in momentum space under an external magnetic field.

Despite the tremendous interest in Weyl semimetals, experimental research devoted to its discovery has been held back due to their rareness in nature. The transition metal monoarsenides have recently been suggested as Weyl semimetals \cite{Hsin_TaAs, Dai_TaAs, TaAs_Exp_Us, TaAs_Exp_XiDai}. By combining low-energy (vacuum ultraviolet) and soft X-ray angle-resolved photoemission spectroscopies (ARPES), we directly observe both the Weyl cones in the bulk and the Fermi arc surface states on the (001) cleaving surface of NbAs. Specifically, our soft X-ray data clearly reveals that the bulk bands touch in discrete points in the bulk Brillouin zone (BZ) and disperse linearly along both the in-plane and out-of-plane directions, which, therefore, form the Weyl cones. Our low-energy ARPES measurements reveal the existence of Fermi arc surface states and show that they connect the projected bulk Weyl nodes. Our systematic ARPES data, and its excellent agreement with our previous theoretical prediction and first-principle band structure calculations, collectively demonstrate the Weyl semimetal state in NbAs. Our discovery not only paves the way for the many fascinating topological quantum phenomena predicted in Weyl semimetal state \cite{Chiral,Aji2012,Nonlocal,Carbotte2013,Franz2013, Ojanen}, but also establishes a new keystone in the correspondence and synergy between high-energy and condensed matter physics.

Niobium arsenide, NbAs, crystallizes in a body-centered tetragonal Bravais lattice, space group $\textit{I}4_1md$ (109), point group $C_{4v}$. Our X-ray diffraction (XRD) obtains lattice constants of $a = 3.45$ $\textrm{\AA}$ and $c = 11.68$ $\textrm{\AA}$, consistent with two earlier crystallographic studies \cite{NbAs_Crystal_1, NbAs_Crystal_2}. The basis consists of two Nb atoms and two As atoms (Fig.~\ref{Fig1}\textbf{a}). The lattice can be understood as a stack of square lattice layers, where each layer consists entirely of either Nb or As atoms. In the crystal, each layer is shifted by half a lattice constant in either the $x$ or $y$ direction relative to the layer below it. These shifts give the lattice a screw pattern along the $z$ direction, which leads to a non-symmorphic $C_4$ rotation symmetry that requires a translation along the $z$ direction by $c/4$, or one-fourth a lattice constant in the conventional unit cell. More importantly, it can be seen that the NbAs crystal lacks space inversion symmetry. Since the system respects time-reversal symmetry, the broken inversion symmetry is a crucial condition, without which the Weyl semimetal phase is not possible. Fig.~\ref{Fig1}\textbf{b} shows a scanning tunneling microscope image on the cleaved (001) surface. The topography image clearly resolves a square lattice without any obvious defect, demonstrating the high quality of our sample. From the topography, we obtain a lattice constant $a = 3.52$ $\textrm{\AA}$, consistent with the value determined by XRD.

We present an overview of the band structure of NbAs before discussing our ARPES results. Fig.~\ref{Fig1}\textbf{c} shows the calculated band structure without spin-orbit coupling. It can be seen that the conduction and valence bands cross each other near the $\Sigma$, $\Sigma'$ and $N$ points, in agreement with the semimetal groundstate. In order to understand the detailed configuration of the band crossing in momentum space, we have calculated the band structure throughout the bulk BZ. As shown in the left panel of Fig.~\ref{Fig1}\textbf{d}, without spin-orbit coupling, the conduction and valence bands dip into each other, giving rise to four 1D ring-like crossings at the $k_x=0$ and $k_y=0$ planes. After the inclusion of spin-orbit coupling (The left panel of Fig.~\ref{Fig1}\textbf{d}), the bands become gapped everywhere along the rings, but a number of point-like band crossings, the Weyl nodes, emerge in the vicinity of the rings. Our calculation shows that there are in total 24 Weyl nodes, which can be categorized into two groups. We name the 8 Weyl nodes that are located on the $k_z=2\pi/c$ plane as W1, and the other 16 away from the $k_z=2\pi/c$ plane as W2. We further study the projection of these Weyl nodes onto the (001) surface. All W1 Weyl nodes project as single Weyl nodes in the close vicinity of the surface BZ edge $\bar{X}$ and $\bar{Y}$ points. On the other hand, interestingly, each pair of W2 Weyl nodes with the same chiral charge are projected onto the same location on the (001) surface BZ. Therefore, there are 8 projected W2 Weyl nodes with the projected chiral charge of $\pm2$. The distribution of the projected Weyl nodes and their chiral charges are shown in the right panel of Fig.~\ref{Fig1}\textbf{d}. The theoretically calculated and ARPES (001) Fermi surface measured at the incident photon energy of 50 eV are shown in Figs.~\ref{Fig1}\textbf{e}, respectively. A good agreement is found between the ARPES and the calculation results. We discuss some general features of the Fermi surface before going into details. The surface BZ is found to be a square. Moreover, the momentum space distance between the surface BZ center $\bar{\Gamma}$ and the edge $\bar{X}$ or $\bar{Y}$ is about $0.91$ $\textrm{\AA}^{-1}$ in our ARPES data. These observations are in agreement with the tetragonal lattice ($a=b=3.45$ $\textrm{\AA}$) of NbAs and the lattice constant ($\pi/a=\pi/3.45=0.91$). Both our ARPES data and calculation show that the Fermi surfaces are located near the surface BZ edge $\bar{X}$ and $\bar{Y}$ points and the midpoints between the $\bar{\Gamma}-\bar{X}$ and the $\bar{\Gamma}-\bar{Y}$ lines. Furthermore, it can be seen from our data and calculation that the Fermi surface violates the $C_4$ rotational symmetry, even though the surface BZ is a square. This is entirely consistent with the crystal structure, where the rotational symmetry is implemented as a screw axis that sends the crystal back into itself after a $C_4$ rotation and a translation by $(0,\frac{a}{2},\frac{c}{4})$ along the rotation axis. As a result, the (001) surface breaks the rotational symmetry of the crystal and the surface Fermi surface does not need to respect the $C_4$ symmetry.

We now present our ARPES data to demonstrate the existence of the bulk Weyl cones and the surface Fermi arcs in NbAs. We have systematically studied the electronic structure of NbAs using both low-energy (vacuum ultraviolet, 35 eV to 90 eV in our experiments) and soft X-ray (350 eV to 1000 eV in our experiments) ARPES. We found that at the vacuum ultraviolet incident photon energies (e.g. Fig.~\ref{Fig1}\textbf{e}), the surface states dominate the ARPES spectral weight, whereas the bulk bands are very weak. On the other hand, at soft X-ray energies the situation is reversed and, therefore, the bulk bands become predominant, which is consistent with the enhanced bulk sensitivity at the soft X-ray energies.

We start by presenting the measurement of the bulk Weyl cones using soft X-ray ARPES. Fig.~\ref{Fig2}\textbf{a} shows the calculated $k_z-k_x$ Fermi surface contour of the bulk bands at the $k_y=0$ plane. The Fermi surface consists of identical contours centered the $\Sigma$ and $S$ points. We note that these contours arise from the 1D nodal line crossings at the $k_y=0$ plane without spin-orbit coupling (Fig.~\ref{Fig1}\textbf{d}), which explains why their shape is very similar to the nodal line crossings. As discussed above in Fig.~\ref{Fig1}\textbf{d}, we do not expect to observe the Weyl nodes in Fig.~\ref{Fig2}\textbf{a} because all Weyl nodes are away from the $k_y=0$ plane. Specifically, W1 nodes are located at the $k_y=0.005$ $\pi/a$ plane whereas W2 nodes are at the $k_y=0.012$ $\pi/a$ plane according to our calculation. Fig.~\ref{Fig2}\textbf{b} shows the ARPES measured $k_z-k_x$ Fermi surface over multiple BZs along the $k_z$ direction. We find a good agreement between our ARPES data and the calculation. Furthermore, we emphasize that the clear dispersion along the $k_z$ direction observed in Fig.~\ref{Fig2}\textbf{b} firmly shows that our soft X-ray ARPES primarily measures the bulk bands instead of the surface states. This is accurate because if we were measuring the surface states, then one would expect the bands to be $k_z$ independent, which means that the Fermi surface consists of straight lines that are parallel to the $k_z$ axis.

In our calculation, the location of the W2 is $(k_x,k_y,k_z)=(0.561\pi/a, 0.012\pi/a, 1.16\pi/c)$. We use that as a guideline, and choose an incident photon energy (i.e. a $k_z$ value) that corresponds to the W2 Weyl nodes. Fig.~\ref{Fig2}\textbf{d} shows the ARPES measured $k_x-k_y$ Fermi surface at the $k_z$ value that corresponds to the W2 Weyl nodes using incident photon energy of 651 eV. We compare the soft X-ray and low energy Fermi surface Fig.~\ref{Fig2}\textbf{d} and Fig.~\ref{Fig1}\textbf{e}. Although they are measured on the same (001) surface of the same sample, the soft X-ray ARPES Fermi surface contains only Fermi points and is $C_4$ symmetric and, whereas, the low energy ARPES Fermi surface contains many pockets and violates the $C_4$ symmetry. This, again, provides compelling evidence for the soft X-ray ARPES measuring the bulk bands while low energy ARPES is sensitive to the surface states. We show the schematic of the Fermi surface at the $k_z$ that corresponds to W2 nodes in Fig.~\ref{Fig2}\textbf{e}. Apart from the eight W2 Weyl nodes, there are additional trivial pockets along the $k_x$ and $k_y$ axes, whcih are denoted by the red circles. This is consistent with our ARPES data in Fig.~\ref{Fig2}\textbf{d}. We therefore study the energy dispersion of the W2 Weyl cones by choosing a momentum space cut direction, Cut 1, which goes across a pair of W2 Weyl nodes, as noted in Fig.~\ref{Fig2}\textbf{d}. The dispersion map is shown in Fig.~\ref{Fig2}\textbf{f} over a wide energy range, where a linear dispersion is clearly observed. Since Cut 1 goes through two nearby W2 Weyl nodes, one would expect to observe two linearly dispersing cones in the ARPES spectra. However, the two cones are very close to each other in momentum space. According to our calculation, the $k$ space distance between the two W2 nodes is $0.024$ $\pi/a\simeq0.022$ $\textrm{\AA}^{-1}$, which is challenging to resolve with soft X-ray ARPES. However, due to the sharp ARPES spectrum collected from our high quality samples, we are able to barely see that the ARPES dispersion in Fig.~\ref{Fig2}\textbf{g} are consistent with the scenario with two cones. We further perform a second derivative image of the dispersion in Fig.~\ref{Fig2}\textbf{h}, where the two cones are better seen. The $k_y$ separation of the two Weyl nodes is about $0.025$ $\textrm{\AA}^{-1}$, which agrees with the calculation. Since a Weyl cone features linear dispersion along all three directions, it is therefore also important to study the dispersion of the W2 Weyl bands along the out-of-plane $k_z$ direction. We fix the $k_x$ and $k_y$ at the location that corresponds to a W2 Weyl node as shown by Cut 2 in Fig.~\ref{Fig2}\textbf{d} and study the energy dispersion as a function of $k_z$. As shown in Fig.~\ref{Fig2}\textbf{i}, the Weyl band is found to be strongly $k_z$ dispersive, which demonstrates its 3D nature. Moreover, throughout the $k_z$ range shown in Fig.~\ref{Fig2}\textbf{i}, the band only crosses the Fermi level at two discrete $k_z$, where the $k_z$ values of these two points are approximately $(52\pm1.2)\pi/c=\pm1.2\pi/c$. This is again consistent with the $k_z$ value of the W2 Weyl nodes from calculation, $k_z=\pm1.16\pi/c$. In the vicinity of the two $k_z$ values where the bands cross the Fermi level, the bands disperse linearly within our instrumental $k_z$ resolution (Fig.~\ref{Fig2}\textbf{i}). Our ARPES measured $E-k_z$ dispersion data in Fig.~\ref{Fig2}\textbf{i} is well consistent with the corresponding theoretical calculation in Fig.~\ref{Fig2}\textbf{j}.

Next, we use the calculated momentum space location of Weyl nodes W1, $(k_x,k_y,k_z)=(0.976\pi/a, 0.004\pi/a, 0\pi/c)$ as a guideline and choose a photon energy that corresponds to W1. Fig.~\ref{Fig2}\textbf{k} shows the $k_x-k_y$ Fermi surface at the $k_z$ value that corresponds to W1. The schematics of the expected Fermi surface at  $k_z=$W1 is shown in Fig.~\ref{Fig2}\textbf{l}. It can be seen that the Fermi surface at $k_z=$W1 consists of both the W1 Weyl nodes that are very close to the surface BZ edge [$(k_x, k_y)=(\pi/a, 0)$ or $(0,\pi/a)$] and some trivial pockets closer to the BZ center. Therefore, we study the energy dispersion of the W1 Weyl cone by choosing a momentum space cut direction, Cut 3, as noted in Fig.~\ref{Fig2}\textbf{k}. We observe a clear dispersion of the W1 Weyl cone in Figs.~\ref{Fig2}\textbf{m,n}. Since Cut 3 goes across two nearby W1 Weyl nodes. One should, in principle, observe a pair of Weyl cones in the dispersion map depicted by Figs.~\ref{Fig2}\textbf{m,n}. However, because the two W1 Weyl nodes are extremely close to each other (the $k$ space separation is 0.007 $\textrm{\AA}^{-1}$ in our calculation, which is much smaller than in W2), our experimental resolution is not sufficient. Therefore, the separation of the two cones for W1 cannot be resolved. Moreover, our data shows that the Fermi level is about 38 meV above the energy of the W1 Weyl node and, as a result, a small part of the upper Weyl cone is seen. This is in agreement with the results from our first principle calculation, shown in Fig.~\ref{Fig2}\textbf{c}. More precisely, our calculation shows that the energy difference between the W1 and W2 Weyl nodes is found to be 36 meV.  Consequently, the systematic measurements presented here, when corroborated by our calculations, experimentally reveal the bulk Weyl cones in NbAs.

After establishing the existence of the bulk Weyl cones, we study the Fermi arc surface states in NbAs by using low energy (vacuum ultraviolet) ARPES in Fig.~\ref{Fig3}. Fig.~\ref{Fig3}\textbf{a} shows the constant energy contours of the (001) surface at different binding energies with the incident photon energy fixed at 50 eV. It can be seen that the constant energy contours, especially the pockets centered at the BZ edge $\bar{X}$ and $\bar{Y}$ points, violate the $C_4$ symmetry, which serves as the signature for these contours arising from the surface states. At the Fermi level ($E_{\textrm{B}}=0$ meV), our data shows that the Fermi surface consists of three types of pockets: a dog-bone shaped contour centered at the $\bar{Y}$ point, a long elliptical contour centered at the $\bar{X}$ point, and a tadpole-shaped pocket that goes along each the $\Gamma-\bar{X}$ or $\Gamma-\bar{Y}$) line from the midpoint of the line toward the $\bar{X}$ ($\bar{Y}$) point. Our ARPES data and calculation (Fig.~\ref{Fig1}\textbf{e}) show agreement on the tadpole and the dog-bone surface states, but not on the elliptical contour. Note, the calculation result can be tuned by adjusting the surface potential and other surface parameters to further improve its consistency with the data. We focus on the tadpole surface states where the agreement between the data and the calculations is the best. As shown in Fig.~\ref{Fig3}\textbf{e}, the tadpole surface states give rise to the Fermi arcs that are connected to the W2 Weyl nodes. Specifically, the pair of W2 Weyl nodes are located at the joint between the ``head'' and the ``tail'' of the tadpole. Therefore, each W2 is connected to two Fermi arcs, consistent with its projected chiral charge of $\pm2$. Because we have resolved both the W2 Weyl nodes using soft X-ray ARPES (Fig.~\ref{Fig2}\textbf{d}) and the tadpole surface states using low energy ARPES (Fig.~\ref{Fig3}\textbf{a}), we, therefore, superimpose these two Fermi surfaces together to scale, as shown in Fig.~\ref{Fig3}\textbf{b}. Indeed, we find that the W2 nodes are located between the ``head'' and the ''tail'' of the tadpole. This demonstrates that the observed tadpole surface states consists of three Fermi arcs, namely a curve that directly connects the pair of W2 nodes, and two nearly straight lines that runs from each W2 toward the surface BZ edge. On the other hand, resolving the Fermi arc connectivity associated with the W1 Weyl nodes is much more challenging because the W1 nodes are too close to each other (the $k$ space separation is 0.007 $\textrm{\AA}^{-1}$). In fact, it is even difficult to figure out the connectivity associated with the W1 in calculation, because the W1 nodes are too close to each other and because there are additional bulk bands at the $k$-space between the pair of W1 nodes. Nevertheless, our observation of the tadpole Fermi arc surface states that connect the W2 Weyl nodes is sufficient to experimentally establish the existence of Fermi arc surface states in NbAs.

Our direct experimental observation by ARPES of the unusual way in which Fermi arcs connect Weyl points on the (001) surface of NbAs places strong constraints on the dispersion of the Fermi arcs. To understand how, we recall first that the Weyl points W2 are arranged in such a way that, for an NbAs sample cleaved on the (001) surface, two Weyl points of the same chirality project onto the same point of the surface BZ. As a result, each of the Weyl point projections has a chiral charge of $\pm 2$. This chirality requires not only that two Fermi arcs connect to each of the Weyl point projections in the surface BZ, but also that the Fermi arcs connecting to a given Weyl point projection be co-propagating. To clarify this requirement, consider a circular pipe passing through the bulk BZ of NbAs which encloses two Weyl points of chiral charge $+1$, as shown in the bottom panel of Fig. 4a. The Chern number on this manifold is $+2$. By cleaving the NbAs sample on the (001) surface, we introduce a boundary on this manifold corresponding to a circle in the surface BZ, as shown in the middle panel of Fig. 4a. On this circle, the Chern number of $+2$ requires that the band structure have two gapless co-propagating chiral boundary modes, illustrated by the red and blue curves in the top panel of Fig. 4a. By determining the chirality of the Fermi arc surface states using this circle in the surface BZ, we see that if the Weyl points are connected by tadpole Fermi arcs, as we observe in our ARPES spectra, then the Fermi arcs must disperse as illustrated in Fig. 4b. The green cones correspond to the Weyl cones projected onto the surface BZ, while the red and blue sheets correspond to the tadpole Fermi arcs. We can better visualize the evolution as a function of binding energy by considering cuts through this band structure. We illustrate in Fig. 4c constant-energy contours of this band structure at a cut above the energy of the Weyl points $\varepsilon > \varepsilon_W$, near the energy of the Weyl points $\varepsilon \sim \varepsilon_W$ and below the energy of the Weyl points $\varepsilon < \varepsilon_W$. We find that the condition for co-propagating chiral modes actually requires the red surface states to grow with binding energy, while the blue surface states shrink with binding energy. Our prediction of this structure of the tadpole Fermi arc surface states is consistent with our numerical calculation of the chiral charge of Weyl points in NbAs and our direct experimental observation by ARPES of the unusual way in which Fermi arcs connect Weyl points on the (001) surface of NbAs.


\section{Acknowledgement}

MZH, SYX and IB acknowledge theoretical discussions with I. Klebanov, A. Polyakov, P. Steinhardt and H. Verlinde. We gratefully thank Sung-kwan Mo, Jonathan Denlinger, for beamline support at the ALS beamlines 10.0.1 and 4.0.3, respectively. We acknowledge C.-H. Hsu for technical assistance with some parts of the theoretical calculations presented here.
Grants:

\newpage

\begin{figure}
\centering
\includegraphics[width=16cm]{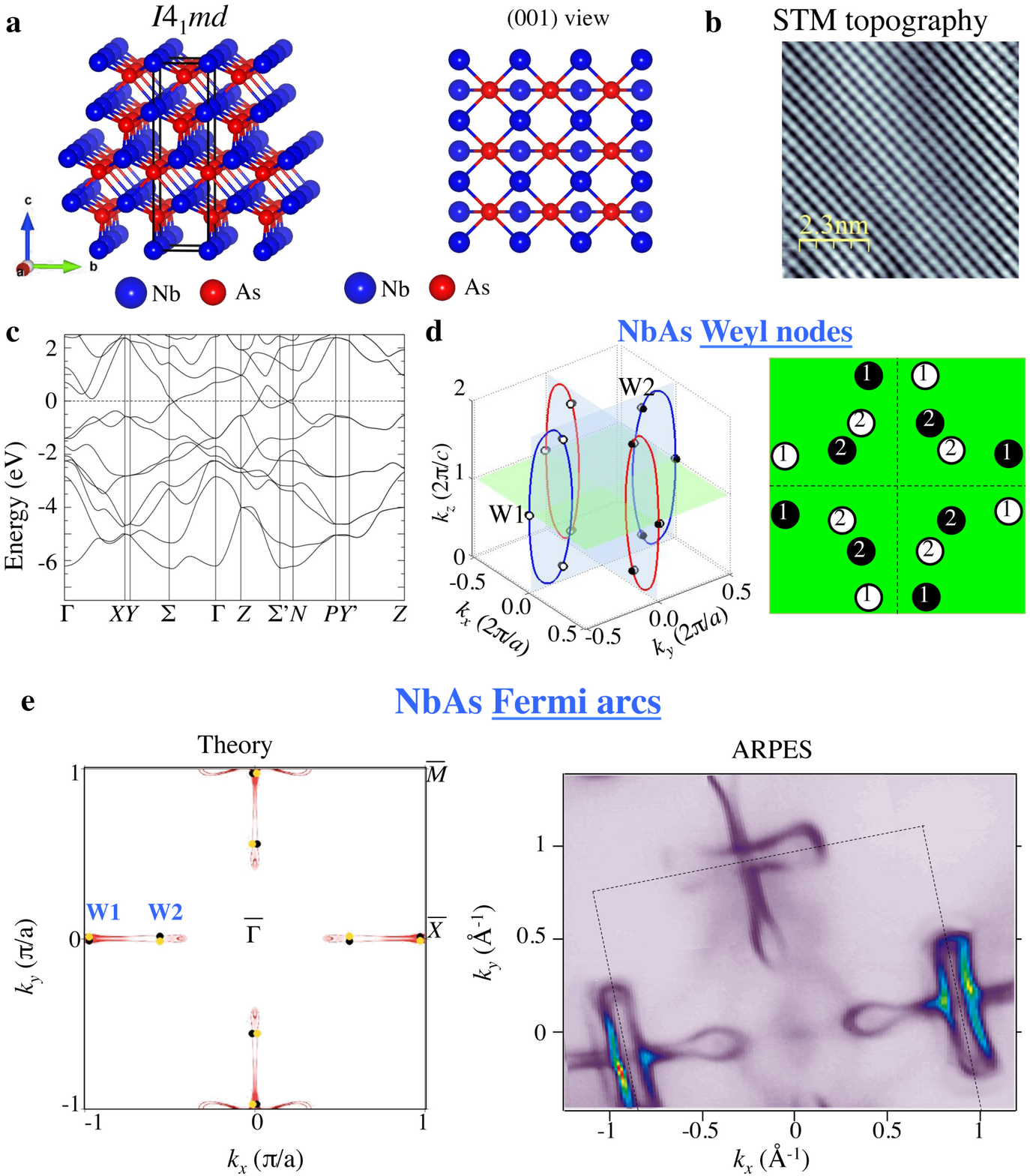}
\caption{\label{Fig1}\textbf{Topological electronic structure of NbAs: Weyl nodes and Fermi arcs.} \textbf{a,} Body-centred tetragonal structure of NbAs, shown as stacks of Nb and As layers. \textbf{b,} Scanning tunneling microscopy (STM) topographic images of cleaved surfaces. \textbf{c,} First-principles band structure calculation of the bulk NbAs without spin-orbit coupling. \textbf{d,} (Left) schematics of the distribution of the Weyl nodes in the three-dimensional Brillouin zone (BZ) of NbAs. The red and blue lines represent the nodal lines without considering spin-orbit coupling. (Right) schematics showing the projected Weyl nodes and their chiral charges on the}
\end{figure}
\addtocounter{figure}{-1}
\begin{figure*}[t!]
\caption{(001) Fermi surface of NbAs. The projected Weyl nodes are denoted by black and white circles whose color indicates the opposite chiral charges of the Weyl nodes. \textbf{e,} First-principles band structure calculated (left) and the ARPES measured Fermi surface (right) of the (001) Fermi surface of NbAs. The Fermi arcs are clearly resolved in our ARPES measurements in agreement with the theoretical prediction.}
\end{figure*}

\clearpage

\begin{figure}
\centering
\includegraphics[width=17cm]{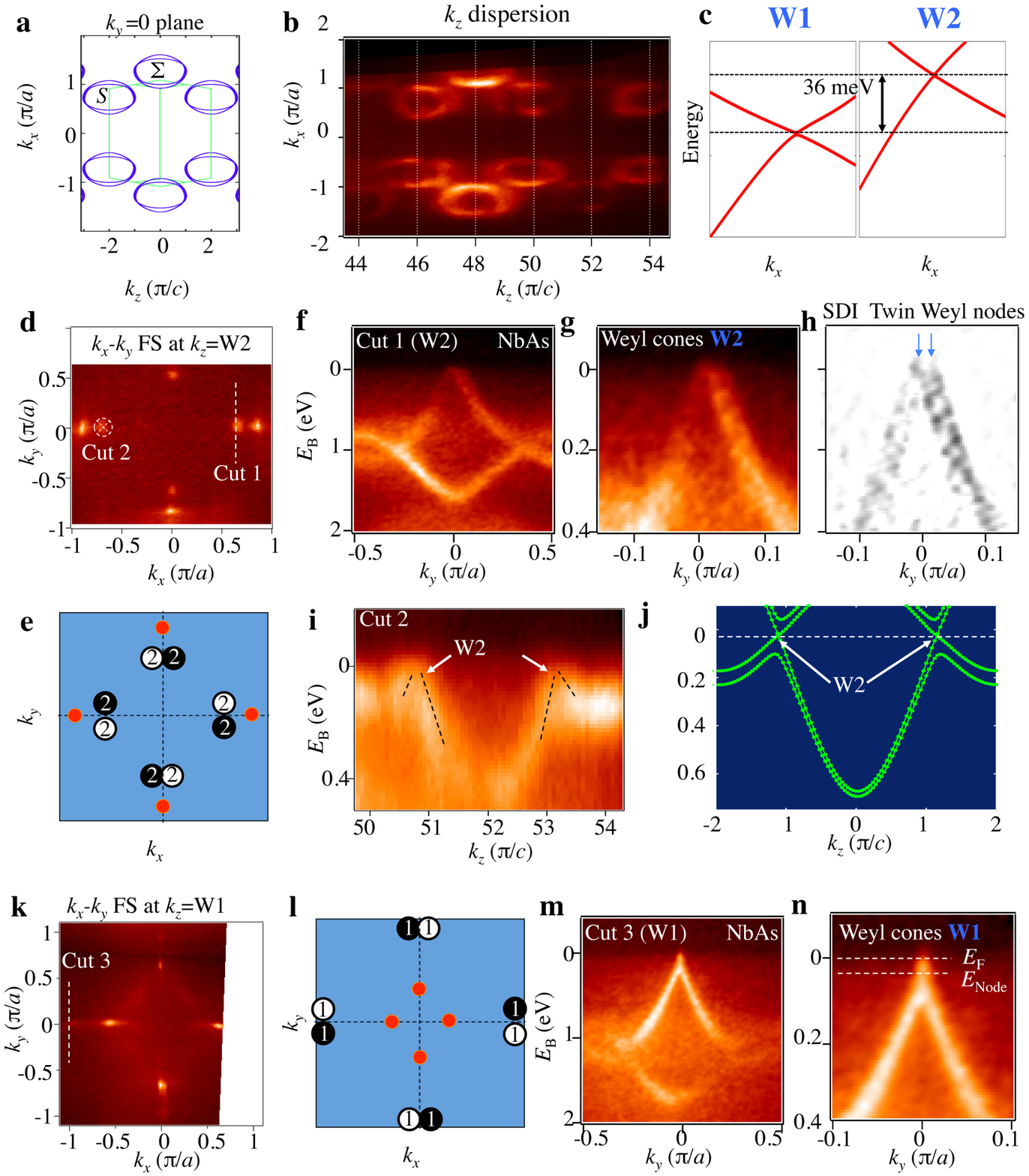}
\caption{\label{Fig2}\textbf{Weyl cones in NbAs.} \textbf{a,} First-principles calculated Fermi surface map at $k_y = 0$ in the $k_z$-$k_x$ plane. \textbf{b,} Soft X-ray ARPES Fermi surface map at $k_y = 0$ in the $k_z$-$k_x$ plane, which agrees well with the theoretical prediction in \textbf{a}. The measurements were conducted}
\end{figure}
\addtocounter{figure}{-1}
\begin{figure*}[t!]
\caption{using photon energies ($\propto$ $k_z$) of 518 - 800 eV. \textbf{c,} First-principles calculations of the energy-momentum dispersions of the two types of Weyl nodes (W1 and W2) in NbAs which show that these nodes are offset in energy by 36 meV relative to each other. \textbf{d,} Soft X-ray ARPES Fermi surface map at $k_z =$ W2 in the $k_x$-$k_y$ plane, revealing the locations of the W2 Weyl nodes. \textbf{e,} Schematics corresponding to \textbf{d} indicating the locations of the W2 Weyl nodes (black and white circles marked as ``2") and other trivial bulk bands (red dots). \textbf{f,} Soft  X-ray ARPES spectra, \textbf{g,} its zoomed-in version close to the Fermi level and \textbf{h,} its curvature plot along Cut 1 direction shown in \textbf{d} that goes through the twin Weyl nodes marked as ``2" in \textbf{e}. The photon energy of the measurements in \textbf{d}, \textbf{f}, and \textbf{g} is 651 eV. \textbf{i,} Soft  X-ray ARPES spectra along Cut 2 direction shown in \textbf{d} (normal to the plane), which again reveals the existence of two Weyl cones along the $k_z$ direction. \textbf{j,} First-principles calculations of the energy-momentum dispersions of the W2 Weyl nodes along $k_z$ corresponding to the ARPES measurements in \textbf{i}. \textbf{k,} Soft X-ray ARPES Fermi surface map at $k_z =$ W1 in the $k_x$-$k_y$ plane, revealing the locations of the W1 Weyl nodes. The photon energy here is 611 eV. \textbf{l,} Schematics corresponding to \textbf{k} indicating the locations of the W1 Weyl nodes (black and white circles marked as ``1") and other trivial bulk bands (red dots). \textbf{m,} Soft  X-ray ARPES spectra and \textbf{n,} its zoomed-in version close to the Fermi level  along Cut 3 direction shown in \textbf{k} that goes through the Weyl nodes marked as ``1" in \textbf{l}. As can be seen in \textbf{n} the energy of the W1 Weyl node is below the Fermi level and W2, in agreement with the calculations in \textbf{c}.}
\end{figure*}

\clearpage

\begin{figure}
\centering
\includegraphics[width=17cm]{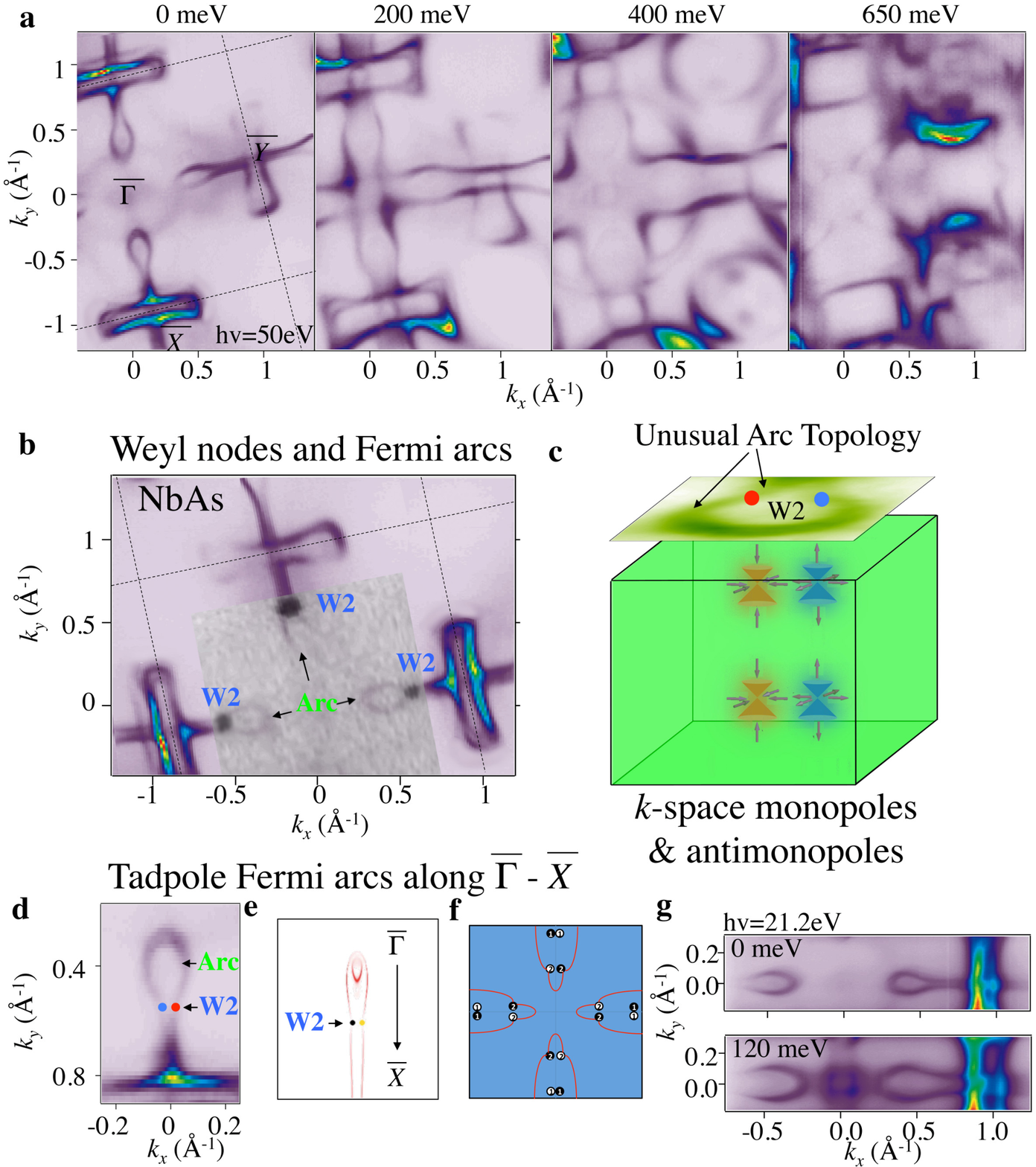}
\caption{\label{Fig3}\textbf{Observation of Fermi arc surface states on the (001) surface of NbAs.} \textbf{a,} High-resolution ARPES Fermi surface map and constant binding energy contours of the band structure of NbAs along the (001) direction at various energies. The square BZ and $C_4$ rotational symmetry at higher binding energies both clearly indicate that the sample is cleaved on the (001) surface. The $C_4$ violation by certain bands at shallow binding energies is consistent with $C_4$ screw axis symmetry broken by the (001) surface and clearly shows that the $C_4$ asymmetric states are}
\end{figure}
\addtocounter{figure}{-1}
\begin{figure*}[t!]
\caption{surface states. \textbf{b,} High-resolution ARPES Fermi surface map, with the soft X-ray ARPES map from 2\textbf{c} overlaid on top of it to scale (grayscale region), showing the relative positions of the Fermi arcs and the Weyl nodes. \textbf{c,} Schematics of the locations of the W2 Weyl nodes in the bulk BZ of NbAs and the unusual topology of the Fermi arcs on the (001) surface Fermi surface in this compound. The pseudo-spin texture near the Weyl nodes with positive and negative chiral charges resembles the magnetic field around magnetic monopoles and antimonopoles. \textbf{d,} ARPES Fermi surface map of the tadpole Fermi arcs along $\bar{\Gamma}-\bar{X}$ and the corresponding Weyl points. \textbf{e,} First-principles calculated Fermi surface map of the tadpole Fermi arcs along $\bar{\Gamma}-\bar{X}$ which shows the qualitative agreement with the ARPES data. \textbf{f,} Schematics of the connectivity of the Fermi arcs in the first BZ. \textbf{g,} ARPES Fermi surface, and a constant binding energy contour at binding energy of 120 meV, of the Fermi arcs along $\bar{\Gamma}-\bar{X}$.}
\end{figure*}

\clearpage
\begin{figure}
\centering
\includegraphics[width=17cm]{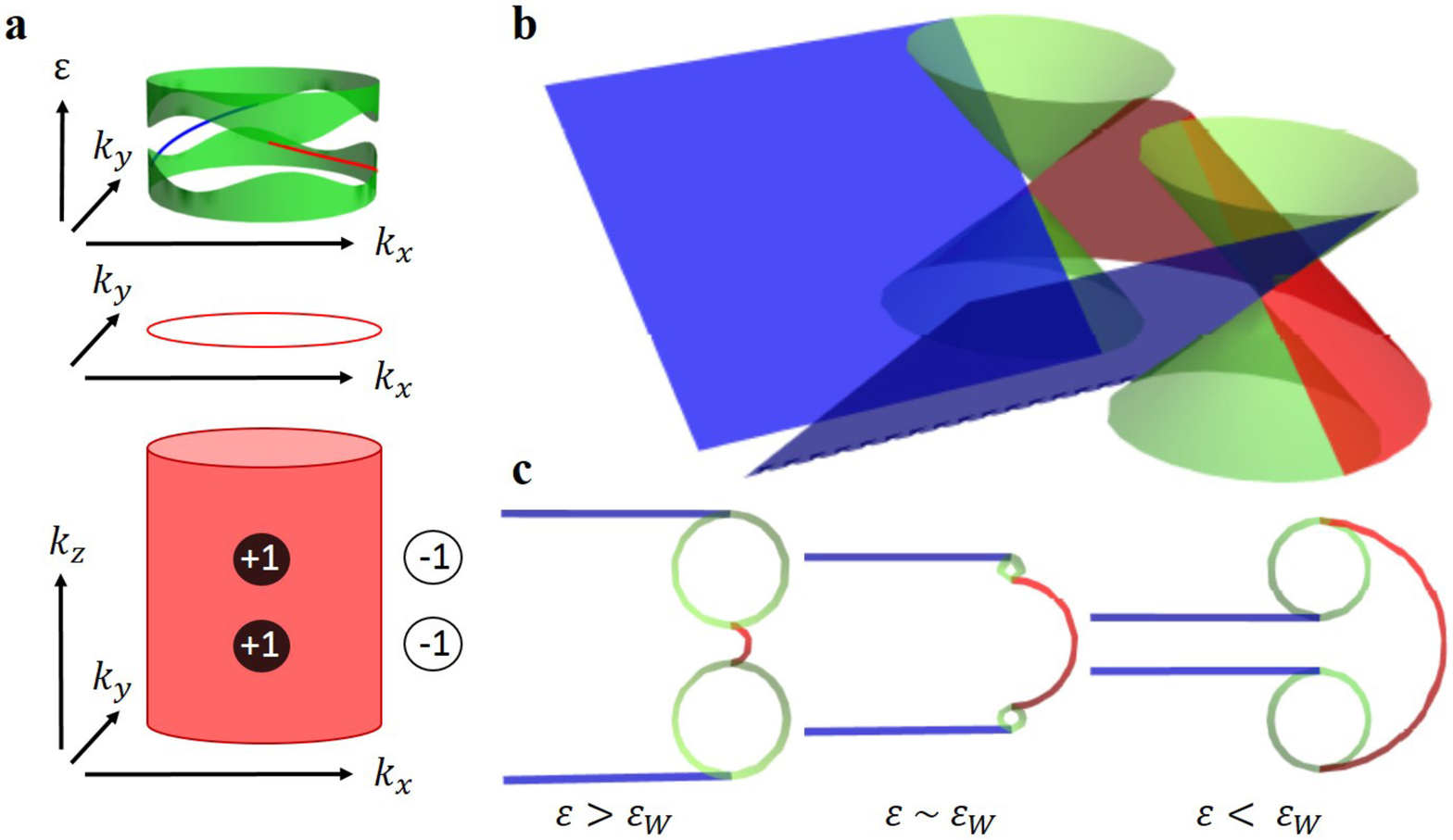}
\caption{\label{Fig4} \textbf{Visualization of the co-propagating chiral modes and the structure of the tadpole Fermi arcs in NbAs.} \textbf{a,} Bottom panel: a circular pipe through the bulk BZ of NbAs, enclosing two Weyl points of chiral charge +1. The pipe extends through the entire bulk BZ in the $k_z$ direction, forming a closed manifold. Because it encloses two Weyl points of chiral charge +1, the Chern number on this manifold is +2. Middle panel: by cleaving an NbAs sample on the (001) surface, we introduce a boundary on this manifold corresponding to a circle in the surface BZ. Top panel: the Chern number of +2 requires that the band structure on this circle in the surface BZ have two gapless co-propagating chiral edge modes. \textbf{b,} Our direct experimental observation by ARPES of the way in which Fermi arcs connect Weyl points on the (001) surface of NbAs places strong constraints on the dispersion of the Fermi arcs. To be consistent with a chiral charge of $\pm 2$ for the projections of the Weyl points on the surface BZ, the tadpole Fermi arcs must disperse as shown by the red and blue sheets. The Weyl cones are shown in green. \textbf{c,} Constant-energy contours of the tadpole Fermi arc surface states at binding energies $\varepsilon$ above the energy of the Weyl points $\varepsilon_W$, near the energy of the Weyl points and below the energy of the Weyl points. For the Fermi arcs to be co-propagating, the red sheet must grow with binding energy, while the blue sheets must disperse toward each other.}
\end{figure}

\end{document}